\documentclass[12pt,english,preprint]{revtex4-1}
\usepackage{times}
\usepackage{array}
\usepackage{longtable}
\usepackage{float}
\usepackage{amsmath}
\usepackage{bm}
\usepackage{graphicx}
\usepackage{amssymb}
\usepackage{color}
\usepackage[FIGTOPCAP]{subfigure}

\makeatletter
\usepackage{fancyhdr}
\usepackage{babel}
\makeatother
\begin{document}

\title{A physics-constrained deep learning surrogate model of the runaway electron avalanche growth rate}

\author{Jonathan Arnaud}
\email{j.arnaud@ufl.edu}
\affiliation{Nuclear Engineering Program, Department of Materials Science and Engineering, University of Florida, Gainesville, FL 32611, United States of America}

\author{Tyler Mark}

\affiliation{Nuclear Engineering Program, Department of Materials Science and Engineering, University of Florida, Gainesville, FL 32611, United States of America}
\author{Christopher J. McDevitt}
\email{cmcdevitt@ufl.edu}
\affiliation{Nuclear Engineering Program, Department of Materials Science and Engineering, University of Florida, Gainesville, FL 32611, United States of America}

\date{\today}

\begin{abstract}

A surrogate model of the runaway electron avalanche growth rate in a magnetic fusion plasma is developed. This is accomplished by employing a physics-informed neural network (PINN) to learn the parametric solution of the adjoint to the relativistic Fokker-Planck equation. The resulting PINN is able to evaluate the runaway probability function across a broad range of parameters in the absence of any synthetic or experimental data. This surrogate of the adjoint relativistic Fokker-Planck equation is then used to infer the avalanche growth rate as a function of the electric field, synchrotron radiation and effective charge. Predictions of the avalanche PINN are compared against first principle calculations of the avalanche growth rate with excellent agreement observed across a broad range of parameters.

\end{abstract}

\maketitle


\section{Introduction}

The unintentional generation of a large relativistic electron population continues to pose a substantial obstacle to the success of the tokamak reactor concept. Such runaway electrons (RE) may be inadvertently generated by the strong electric fields coinciding with a tokamak disruption and obtain energies of several MeV~\cite{Hender:2007}. Due to their high energy and often localized impact, REs have the potential to induce substantial damage to plasma facing components~\cite{matthews2016melt}. Obtaining a robust description of their formation processes has thus emerged as a topic of immediate importance to tokamak devices in addition to its intrinsic interest to plasma physics.

One of the primary challenges in describing REs is the multi-physics nature of a fusion plasma. In particular, alongside a description of RE kinetics, an accurate treatment of RE formation requires the self-consistent evolution of the background magnetohydrodynamic (MHD) equilibrium, impurity transport, and radiative losses. As a result, a promising approach toward developing an integrated description of RE formation and evolution involves the identification of reduced models of RE kinetics that can be coupled to a broader plasma physics framework~\cite{hoppe2021dream,bandaru2021magnetohydrodynamic, liu2021self, sainterme2024resistive}. Developing such a reduced RE module has, however, posed a challenge to the plasma physics community. While several reduced models of runaway generation processes such as Dreicer generation~\cite{Dreicer:1959, Kruskal-Bernstein:1962, Connor:1975, hesslow2019evaluation}, hot tail generation~\cite{helander2004electron, smith2008hot, yang2023pseudo}, or avalanche generation~\cite{Rosenbluth:1997, Martin:2015, Aleynikov:2015, hesslow2019influence} have been developed, these reduced models often struggle to quantitatively describe the inherently kinetic physics that characterize RE generation. Furthermore, many of these models were derived under highly simplified assumptions, where their generalization to more realistic plasma conditions is non-trivial. 

The rapidly evolving field of deep learning suggests a new pathway to developing reduced RE models. While often computationally intensive to train, the online deployment of deep learning based models is typically orders of magnitude faster than traditional plasma physics codes thus providing an efficient surrogate that may be called by a broader plasma physics framework.
In the present paper, our aim will be to develop a physics-informed neural network (PINN) to provide an accurate reduced model of the RE avalanche. In contrast to purely data driven paradigms, this deep learning approach embeds the physics model into the training of a deep neural network, enabling the PINN to make predictions in the absence of 
synthetic or experimental data. In so doing, the trained PINN encodes the underlying kinetic solution, allowing for greater interpretability, along with predicting quantities of interest (QoI) such as the rate of RE generation.


This approach was recently used to predict the number of hot tail seed electrons in an axisymmetric model of the thermal quench~\cite{McDevitt:hottail:2023}. Our aim in the current paper is to develop a PINN to describe the avalanche mechanism of REs~\cite{Sokolov:1979, Jayakumar:1993}. 
In contrast to the hot tail seed mechanism, the avalanche mechanism of RE formation requires a pre-existing `seed' RE population to be present. Once established, large-angle collisions of this RE seed with the cold background plasma will result in the initially cold electrons being scattered to energies up to half of the initial energy of the seed electron. For a sufficiently large electric field, these `secondary' electrons will be accelerated to relativistic energies resulting in the exponential growth of the original seed. Such a process is of particular importance to reactor scale tokamak plasmas due to the ability of this mechanism to convert the Ohmic plasma current into RE current, even in the limit where only a minuscule seed population of REs is present~\cite{Martin:2017, vallhagen2020runaway}.

The remainder of this paper is organized as follows: Section \ref{sec:PCDL} provides a brief overview of the physics-informed deep learning framework employed. The adjoint of the relativistic Fokker-Planck equation, together with its solution, is described in Sec. \ref{sec:RPF}. Section \ref{sec:PIE} evaluates the avalanche growth rate and threshold across a broad range of parameters and verifies the predictions against those from a traditional RE solver. Conclusions along with a brief discussion are given in Sec. \ref{sec:SC}.

\section{\label{sec:PCDL}Physics-Constrained Deep Learning}
Physics-constrained deep learning methods have emerged as a powerful means to efficiently describe complex physical processes. In addition to exploiting available data, such methods seek to embed physical constraints into the training of a neural network~\cite{lagaris1998artificial, karpatne2017theory, karniadakis2021physics, lusch2018deep, wang2020towards}, thus providing a natural means of avoiding overfitting, along with allowing for greater generalizability to unseen parameter regimes. Physics-informed neural networks~\cite{karniadakis2021physics} have emerged as a particularly prominent example. A PINN, in its simplest form, incorporates the partial differential equation (PDE), boundary, and initial conditions into the loss function, yielding
\begin{align}
\text{Loss} &= \frac{1}{N_{PDE}} \sum^{N_{PDE}}_i  \mathcal{R}^2 \left( \mathbf{p}_i, t_i; \bm{\lambda}_i \right) + \frac{1}{N_{bdy}} \sum^{N_{bdy}}_i \left[ P_i - P \left( \mathbf{p}_i, t_i; \bm{\lambda}_i \right)\right]^2 \nonumber \\
& + \frac{1}{N_{init}} \sum^{N_{init}}_i \left[ P_i - P \left( \mathbf{p}_i, t=0; \bm{\lambda}_i \right)\right]^2
, \label{eq:PCDL2}
\end{align}
where $P \left( \mathbf{p}, t; \bm{\lambda} \right)$ is the dependent variable (the runaway probability function for the present work), $\mathcal{R}\left( \mathbf{p}, t; \bm{\lambda} \right)$ is the residual of the PDE, $\mathbf{p}$ and $t$ are the independent variables (momentum and time), and $\bm{\lambda}$ represents parameters of the physical system.
Here, the first term represents the loss against the PDE, the second term represents the loss against the boundary conditions, and the third term represents the loss against the initial conditions for time-dependent problems. Noting that derivatives of the neural network output $P$ with respect to its inputs ($\mathbf{p}, t; \bm{\lambda} $) can be evaluated by automatic differentiation, a feature provided by standard machine learning libraries~\cite{abadi2016tensorflow, paszke2017automatic}, no discretization of the PDE is required. As a result, PINNs are inherently mesh-free and only require specification of  a distribution of training points. After minimization of the loss function, the dependent variable $P$ will satisfy the PDE, boundary, and initial conditions up to the loss defined by Eq. (\ref{eq:PCDL2}). Hence, a PINN provides a means of solving PDEs, where the value of the loss achieved after training provides an estimate of the accuracy of the solution. 

A powerful property of PINNs is that they may be used to learn the parametric solution to a PDE~\cite{sun2020surrogate, mcdevitt2024physics}. In particular, since the parameters of the physical problem $\bm{\lambda}$ are inputs into the neural network, after minimization of the loss, the PINN can predict $P \left( \mathbf{p}, t; \bm{\lambda} \right)$ across a broad range of parameters $\bm{\lambda}$. While obtaining a parametric solution of a PDE often requires extensive offline training, the online execution of a PINN is rapid, where an individual prediction typically requires a few microseconds. Physics-informed neural networks thus provide a framework for developing efficient surrogate models of a PDE. A significant limitation of the above approach, however, is that PINNs often fail to train when treating the challenging PDEs that characterize many scientific and engineering applications~\cite{wang2022and}. A primary aim of this paper will therefore be to develop custom output layers to the PINN that enable it to robustly evaluate the adjoint to the relativistic Fokker-Planck equation across a broad range of plasma conditions in the absence of synthetic or experimental data.


When carrying out the training of the PINN, a single Nvidia A100 GPU will be used. A fully connected feedforward neural network is employed, containing six hidden layers, each having a width of 64 neurons. Roughly a million training points are used and distributed across the input space $\left( \mathbf{p}, t, \bm{\lambda} \right)$ according to a Hammersley distribution. During training a small number of training points are added to regions where the residual is maximal. An independent set of test points are also applied to verify accuracy of predictions away from training points. The python script used for training the PINN is written using the DeepXDE library~\cite{lu2021deepxde} with TensorFlow~\cite{abadi2016tensorflow} as the backend.

\section{\label{sec:RPF}Steady State Runaway Probability Function}

\subsection{\label{sec:LDCF}Adjoint Relativistic Fokker-Planck Equation}

The adjoint to the relativistic Fokker-Planck equation has been treated in a variety of contexts, including wave-driven currents in magnetized fusion plasmas~\cite{antonsen1982radio, taguchi1983effect, fisch1986transport, karney1986current} and runaway electron formation~\cite{karney1986current, Liu:2016, Liu:2017, zhang2017backward, McDevitt:hottail:2023}. In the absence of synchrotron radiation, the adjoint to the steady state relativistic Fokker-Planck equation can be written as:
\begin{subequations}
\label{eq:LDCF1}
\begin{align}
&\left[ -E_\Vert \xi - C_F \right]\frac{\partial P }{\partial p} - \left( 1-\xi^2\right)\frac{E_\Vert}{p} \frac{\partial P }{\partial \xi} = -\frac{\nu_D}{2} \frac{\partial}{\partial \xi} \left[ \left( 1-\xi^2\right) \frac{\partial P }{\partial \xi}\right]
, \label{eq:LDCF1a}
\end{align}
where the collisional coefficients are taken to be:
\begin{equation}
\nu_D = \left( 1+Z_{eff} \right)\frac{\gamma}{p^3}, \quad C_F = \frac{1 + p^2}{p^2}
. \label{eq:LDCF1b}
\end{equation}
\end{subequations}
Here, the relativistic momentum $p$ is normalized as $p\to p/ \left( m_e c\right)$, the electron's pitch is defined by $\xi \equiv p_\Vert / p$, time is normalized as $t \to t / \tau_c$, where $\tau_c \equiv 4\pi \epsilon^2_0 m^2_e c^3/ \left( e^4 n_e \ln \Lambda \right)$ is the collision time of a relativistic electron, the collisional coefficients $\nu_D$ and $C_F$ are normalized to $\tau_c$, and the parallel electric field is normalized to the Connor-Hastie electric field $E_\Vert \to E_\Vert/E_c$, where $E_c \equiv m_e c / \left( e \tau_c \right)$~\cite{Connor:1975}. Energy diffusion has been neglected due to this term being exceptionally small for conditions typical of a tokamak disruption. In particular, noting that the energy diffusivity scales with $T_e/\left( m_e c^2 \right)$ for the high energies characteristic of REs, we expect this approximation to be well satisfied for a low temperature post thermal quench plasma.
Furthermore, the collision frequencies used in Eq. (\ref{eq:LDCF1}) assume the limit $v>v_{Te}$, where $v_{Te}$ is the electron thermal velocity. For the parameters of interest, the critical speed for an electron to run away will be much larger than $v_{Te}$, hence we anticipate that this approximation will not substantially impact our results.

Concerning the boundary conditions, we will enforce $P=0$ at $p=p_{min}$ and $P=1$ along the upper boundary, where the energy flux $U_p \equiv -E_\Vert \xi - C_F$ is positive. Specifically, the high energy boundary condition enforces that $P$ is unity for values of the pitch $\xi$ where electric field acceleration exceeds collisional drag such that the electron will be accelerated out of the simulation domain. Further noting that for values of the pitch near $\left| \xi \right| \approx 1$, Eq. (\ref{eq:LDCF1}a) is nearly hyperbolic, we anticipate that electrons accelerated through the high energy boundary will be accelerated to arbitrarily high energies, and thus have a zero probability of returning to the simulation domain. Hence, any electron located at the high energy boundary $p_{max}$ with $U_p > 0$, is treated as a RE. With these boundary conditions, the quantity $P \left( p, \xi \right)$ indicates the probability that an electron initially located at $\left( p, \xi \right)$ will run away at a later time, and is often referred to as the runaway probability function (RPF)~\cite{karney1986current}.

When near the threshold electric field for avalanche generation, synchrotron radiation substantially impacts the RPF. While adding synchrotron radiation leads to a modest modification of Eq. (\ref{eq:LDCF1}), it does complicate the physical interpretation of the RPF. In particular, with the inclusion of synchrotron radiation, the adjoint equation takes the form:
\begin{align}
&\left[ -E_\Vert \xi - C_F - \alpha \gamma p \left( 1 - \xi^2 \right) \right]\frac{\partial P }{\partial p} + \left( 1-\xi^2\right) \left[ -\frac{E_\Vert}{p} + \alpha \frac{\xi}{\gamma} \right] \frac{\partial P }{\partial \xi} = -\frac{\nu_D}{2} \frac{\partial}{\partial \xi} \left[ \left( 1-\xi^2\right) \frac{\partial P }{\partial \xi}\right]
, \label{eq:LDCF2}
\end{align}
where the strength of synchrotron radiation is set by the parameter $\alpha \equiv \tau_c/\tau_s$, where $\tau_s \equiv 6 \pi \epsilon_0 m^3_e c^3 / \left( e^4 B^2 \right)$. Here, Eq. (\ref{eq:LDCF2}) is solved similarly as before with a boundary condition of $P=0$ at $p=p_{min}$ along with the condition that $P=1$ on the high energy boundary $p=p_{max}$ when the energy flux, now defined by $U^{(\alpha)}_p \equiv -E_\Vert \xi - C_F - \alpha \gamma p \left( 1 - \xi^2 \right)$, is positive. While this problem formulation is directly analogous to the case that neglects  synchrotron radiation, the physical interpretation of the RPF is slightly modified. Specifically, as shown in Refs. \cite{Andersson:2001, Decker:2016, guo2017phase, mcdevitt2018relation}, synchrotron radiation damping and pitch-angle scattering results in electrons obtaining a saturated energy, achieved via the formation of a circulation pattern in momentum space centered about an O-point. 
Thus, an electron accelerated through the high energy boundary will have a non-zero probability of returning to the simulation domain after a finite time, rather than being accelerated to arbitrarily high energy as was the case when synchrotron radiation was neglected. For the case with synchrotron radiation, the RPF should thus be given the narrower interpretation as the probability that an electron reaches the high energy boundary of the simulation domain before slowing down to the low energy bulk, rather than the probability of an electron being accelerated to arbitrarily high energy.

\subsection{\label{sec:RPFP}Embedding Physical Constraints into the PINN}

Our aim in this section will be to develop a PINN framework capable of robustly representing solutions to Eq. (\ref{eq:LDCF2}) across the three-dimensional parameter space $\left( E_\Vert, Z_{eff}, \alpha \right)$. A key component of our description is enforcing a subset of physical properties of the RPF as hard constraints. In particular, we will (i) enforce the low energy boundary condition $P=0$ at $p=p_{min}$, (ii) constrain the RPF to have a range between zero and one, and (iii) recover the limit that the RPF vanishes when $\left| E_\Vert \right| < 1$. These three constraints are enforced by introducing a customized output layer to the neural network of the form:
\begin{subequations}
\label{eq:RPFP1}
\begin{equation}
P^\prime \left( p, \xi \right) \equiv \frac{1}{2} \left[ 1 + \tanh \left( \frac{E_\Vert - 1}{\Delta E}\right)\right] \left( \frac{p-p_{min}}{p_{max}-p_{min}} \right) P_{NN} \left( p, \xi \right)
, \label{eq:RPFP1a}
\end{equation}
\begin{equation}
P \left( p, \xi \right) \equiv \tanh\left( {P^\prime}^2 \left( p, \xi \right) \right)
. \label{eq:RPFP1b}
\end{equation}
\end{subequations}
Here, $P_{NN}$ is the output of the hidden layers of the neural network, and $\Delta E$ is a hyperparameter whose value should satisfy $\Delta E < 1$, where for all cases in this paper it is taken to be $\Delta E = 0.1$. From Eq. (\ref{eq:RPFP1}) it can be verified that regardless of the value of $P_{NN}$, the predicted RPF $P \left( p, \xi \right)$: (i) vanishes at the low energy boundary, (ii) has a range of zero to one, and (iii) vanishes for $E_\Vert < 1$ (only positive electric fields are considered when training the RPF PINN). We note in passing that the output layer defined by Eq. (\ref{eq:RPFP1}) results in both the value, and the first derivative of $P$ vanishing at $p=p_{min}$. While this latter condition is not required when defining the RPF, it will nevertheless be satisfied as long as the value of $p_{min}$ chosen is below the momentum $p_{crit} \equiv 1/\sqrt{E_\Vert-1}$ where collisional drag exceeds electric field acceleration. Specifically, since energy diffusion is neglected in the present analysis, electrons located below $p_{crit}$ will have zero probability of running away. Thus, for $p < p_{crit}$ the solution of the RPF will be a constant with a magnitude of zero, and thus is consistent with pure Dirichlet and Neumann boundary conditions at $p=p_{min}$

The loss function employed is taken to have the form:
\begin{equation}
\text{Loss} = \frac{1}{N_{PDE}} \sum^{N_{PDE}}_i  \left[\left( \frac{p^2_i}{1+p^2_i}\right)\mathcal{R} \left( p_i, \xi_i ; \bm{\lambda}_i \right) \right]^2 + \frac{1}{N_{bdy}} \sum^{N_{bdy}}_i \left[ 1 - P \left( p_i, \xi_i ; \bm{\lambda}_i \right)\right]^2
, \label{eq:RPFP2}
\end{equation}
where $\mathcal{R} \left( p_i, \xi_i ; \bm{\lambda}_i \right)$ is the residual of Eq. (\ref{eq:LDCF2}) and $\bm{\lambda}$ represents the physics parameters $\left( E_\Vert, Z_{eff}, \alpha \right)$. Here, the first term in Eq. (\ref{eq:RPFP2}) penalizes deviations from the governing PDE given by Eq. (\ref{eq:LDCF2}), where the prefactor $p^2_i/\left( 1+p^2_i\right)$ removes the low energy divergence of the pitch-angle scattering operator. 
The second term in the loss function defined by Eq. (\ref{eq:RPFP2}) enforces the high energy boundary condition, i.e. $P=1$ at $p_{max}$ when $U^{(\alpha)}_p > 0$. In particular, the boundary points $N_{bdy}$, will only be applied at locations that satisfy both $U^{(\alpha)}_p > 0$ and $p=p_{max}$.

\subsection{\label{sec:PDR}Parametric Dependence of the Runaway Probability Function}

\begin{figure}
\begin{centering}
\includegraphics[scale=0.5]{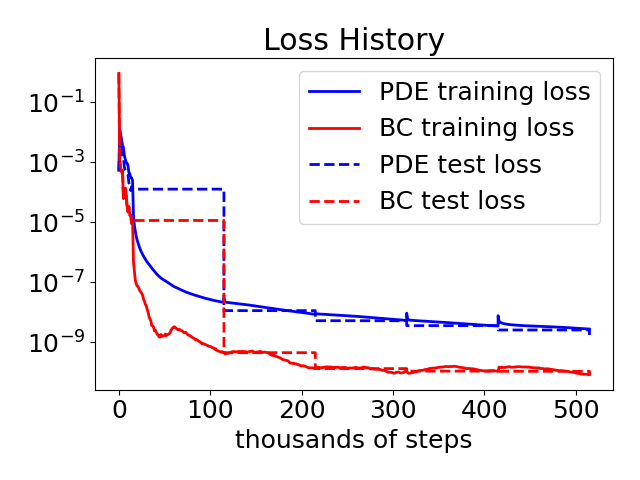}
\par\end{centering}
\caption{Loss history for a feedforward neural network with six hidden layers each with a width of 64 neurons, along with roughly 1,000,000 training points. 15,000 epochs were performed with the ADAM optimizer, with the remaining steps performed using L-BFGS. The model was trained across $E_\Vert \in (1,10)$, $Z_{eff} \in (1,10)$, and $\alpha \in (0, 0.2)$. The range of $p$ was chosen such that the low energy boundary was $10\;\text{keV}$ and the high energy boundary was $5\;\text{MeV}$.}
\label{fig:PDR1}
\end{figure}

In this section, we will seek to obtain solutions to the PINN in the 5D space defined by the two independent coordinates $\left( p, \xi \right)$ and the three physics parameters $\left( E_\Vert, Z_{eff}, \alpha \right)$. The loss history of the PINN is shown in Fig. \ref{fig:PDR1}. Here, after roughly 100,000 epochs, the loss associated with the PDE drops below $10^{-7}$, along with the boundary loss dropping to $\approx 10^{-9}$, with the loss dropping more slowly for the remaining $\sim 400,000$ epochs. The test loss of the PDE reaches $\approx 10^{-9}$ by the end of the training indicating that an accurate solution was found. The periodic spikes in the training data are due to additional  training points sampled after every 100,000 epochs at locations where the residual is maximal. The test loss is only updated after every 100,000 epochs when using the L-BFGS optimizer, leading to the sharp variations evident in the dashed curves.
\begin{figure}
\begin{centering}
\subfigure[]{\includegraphics[scale=0.5]{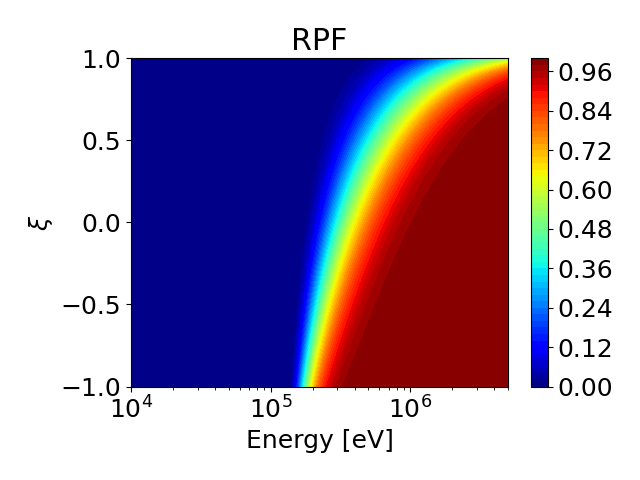}}
\subfigure[]{\includegraphics[scale=0.5]{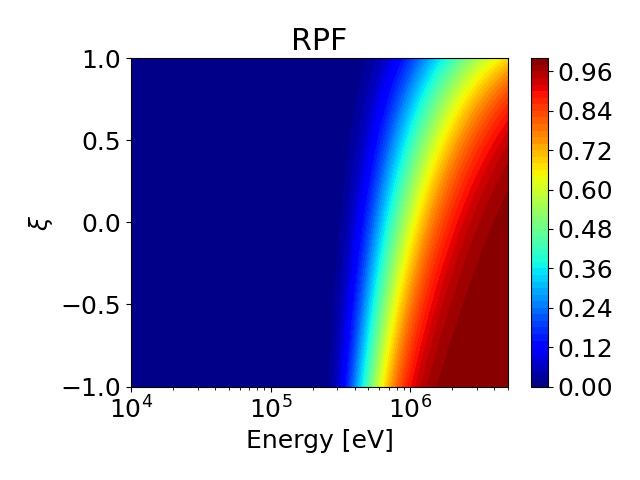}}
\subfigure[]{\includegraphics[scale=0.5]{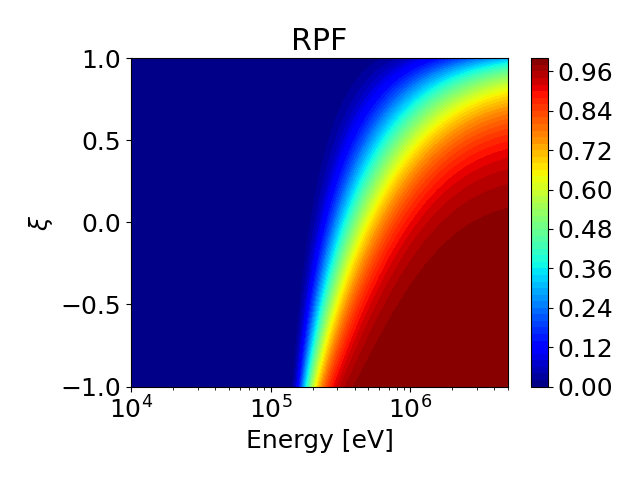}}
\subfigure[]{\includegraphics[scale=0.5]{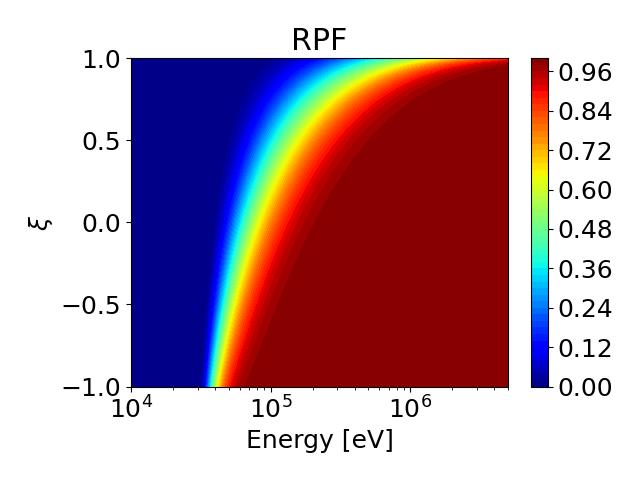}}
\par\end{centering}
\caption{Runaway probability functions for different values of the physics parameters $\left( E_\Vert, Z_{eff}, \alpha \right)$. Panel (a) is for $\left( E_\Vert = 3, Z_{eff} = 1, \alpha = 0\right)$, panel (b) is for $\left( E_\Vert = 3, Z_{eff} = 10, \alpha = 0\right)$, panel (c) is for $\left( E_\Vert = 3, Z_{eff} = 1, \alpha = 0.2\right)$, and panel (d) is for $\left( E_\Vert = 10, Z_{eff} = 1, \alpha = 0\right)$. }
\label{fig:PDR2}
\end{figure}

Four example predictions of the RPF are shown in Fig. \ref{fig:PDR2}. Here, the RPF vanishes at low energies, where drag exceeds electric field acceleration, but increases at higher energy due to the $\left( 1+p^2\right)/p^2$ drop in the collisional drag. In particular, for the parameters indicated in Fig. \ref{fig:PDR2}(a), the $P=0.5$ contour is located at an approximate energy of $200\;\text{keV}$ for $\xi=-1$. Considering a case with a large $Z_{eff}$ [see Fig. \ref{fig:PDR2}(b) with $Z_{eff}=10$], the $P=0.5$ contour shifts to higher energy, with a more gradual transition between the $P\approx 0$ and $P\approx 1$ regions. The impact of synchrotron radiation on the RPF is shown in Fig.  \ref{fig:PDR2}(c), where the magnitude of synchrotron radiation was taken to be $\alpha = 0.2$. Compared to an otherwise identical case, but without synchrotron radiation [Fig. \ref{fig:PDR2}(a)], it is evident that the location of the $P =0.5$ contour at $\xi=-1$ has only shifted slightly. This is due to synchrotron radiation vanishing for $\xi=-1$ and having a modest magnitude at low energies. In contrast, the RPF at high energies is more substantially impacted, where the region with $P \approx 1$ is now largely localized to negative values of pitch for the energy range considered. 
Finally, increasing the electric field to $E_\Vert = 10$, with $Z_{eff}$ = 1 and $\alpha = 0$, results in a drop in the location of the $P=0.5$ contour [compare Figs. \ref{fig:PDR2}(a) and (d)], due to the electric field being able to overcome collisional drag for a larger range of energies.

\subsection{\label{sec:PDR}Critical Energy to Run Away}

\begin{figure}
    \subfigure[]{\includegraphics[scale=0.33]{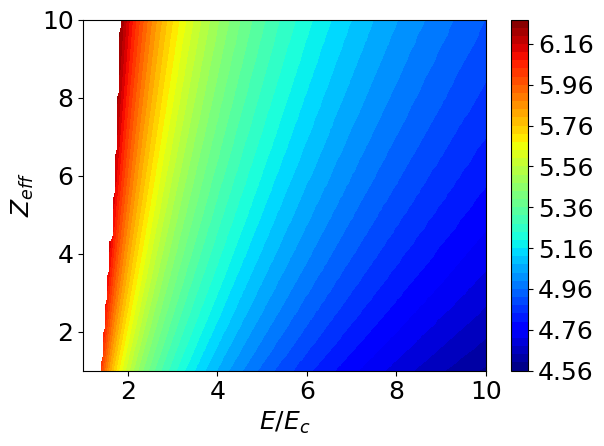}}
    \subfigure[]{\includegraphics[scale=0.33]{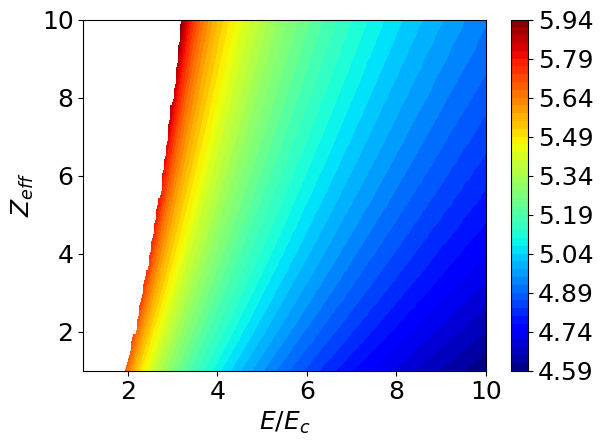}}
    \subfigure[]{\includegraphics[scale=0.33]{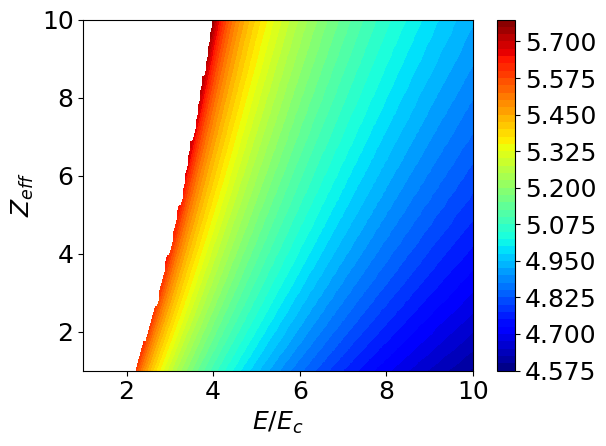}}
    \caption{$\log_{10} E_{crit}$, where $E_{crit}$ is the critical energy to run away in eV. Panel (a) is for $\alpha = 0$, panel (b) is for $\alpha = 0.1$, and panel (c) is for $\alpha = 0.2$.}
    \label{fig:P0o5}
\end{figure}

The location of the $P = 0.5$ contour $\left( p_{crit}, \xi_{crit} \right)$ provides a useful reference point for identifying the critical energy and pitch above which electrons are likely to run away. 
The parametric solution of the PDE given by the PINN thus provides $\left( p_{crit},\xi_{crit} \right)$ across the entire domain by simply extracting the $P = 0.5$ surface. To show the dependence of ($E_\Vert$, $Z_{eff}$, $\alpha$) on the critical energy to run away, we evaluate the energy at which $P = 0.5$ and $\xi = -1$ across the entire training region, which is shown in Fig. \ref{fig:P0o5}. Here, the $\log_{10}$ of the critical energy in units of eV is plotted, where the empty regions represent scenarios where $E_\Vert$ is below the avalanche threshold $E_{av}$ (see Sec. \ref{sec:ARE} below). The dependence of $E_\Vert$ and $Z_{eff}$ on the critical energy can be seen in Fig. \ref{fig:P0o5} (a) for $\alpha = 0$, where increasing the electric field from $E_\Vert = 2$ to $E_\Vert = 10$ decreases the critical energy from $\sim$ 1 MeV to $\sim$ 36 keV. At a given electric field (e.g. $E_\Vert = 4$), the critical energy increases by almost an order of magnitude as $Z_{eff}$ is increased from one to ten, indicating that higher $Z$ elements in the plasma increase the critical energy for REs. Moreover, the addition of synchrotron radiation [compare Figs. \ref{fig:P0o5} (b) and (c)], shows that the region where $E_\Vert < E_{av}$ is increased (larger white region). As $\alpha$ is increased further [compare Figs. \ref{fig:P0o5} (a) and (c)], however, we also see that the shape and range of the critical energy varies modestly, indicating the critical energy is modestly impacted by $\alpha$ when $E_\Vert \gg 1$. 

\section{\label{sec:PIE}Surrogate Model of the Avalanche Growth Rate}

\subsection{\label{sec:ARE}Secondary Source of Runaway Electrons}

Given the rate and distribution of secondary electrons generated by large-angle collisions, the RPF described in Sec. \ref{sec:RPF} can be used to evaluate the avalanche growth rate. In particular, denoting the source of secondary electrons as $S \left( p, \xi \right)$, the rate that REs form due to the avalanche mechanism can be expressed as~\cite{Liu:2017}:
\begin{equation}
\left. \frac{d n_{RE}}{dt} \right|_{av} = \int d^3 p S \left( p, \xi \right) P \left( p, \xi \right)
, \label{eq:PIE1}
\end{equation}
where $n_{RE}$ is the density of REs. Here, $S \left( p, \xi \right)$ indicates the rate and momentum space distribution of REs, whereas $P \left( p, \xi \right)$ indicates the probability that an electron at a given momentum space location $\left( p, \xi \right)$ runs away. By integrating over momentum space, $S \left( p, \xi \right) P \left( p, \xi \right)$ will thus indicate the expected rate that REs form due to the avalanche mechanism. The primary challenge with evaluating Eq. (\ref{eq:PIE1}) is due to $S \left( p, \xi\right)$ depending on the primary RE distribution $f_e$, i.e.
\begin{equation}
S \left( p, \xi \right) = \int d^3 p^\prime S_0 \left(p^\prime , \xi^\prime , p , \xi \right) f_e \left( p^\prime, \xi^\prime \right)
, \label{eq:PIE2a}
\end{equation}
where $S_0 \left(p^\prime , \xi^\prime , p , \xi \right)$ is defined by:
\begin{equation}
S_0 \left(p^\prime , \xi^\prime , p , \xi \right) = n_e c r^2_e \frac{v^\prime}{2\pi p^2} \frac{d \sigma_M \left( p^\prime , p \right)}{dp} \Pi \left( p^\prime, \xi^\prime , p ; \xi \right)
. \label{eq:PIE2b}
\end{equation}
where, $r_e = e^2/\left( 4\pi\epsilon_0 m_e c^2\right)$ is the classical electron radius, $d \sigma_M/dp$ is the M\o ller cross section~\cite{Moller:1932,Ashkin:1954}, $\Pi \left( p^\prime, \xi^\prime , p ; \xi \right)$ describes the pitch-angle dependence of secondary electron generation (see Ref. \cite{Boozer:2015} for an explicit expression), $d^3p = 2\pi p^2 dp d\xi$, and all variables have been dedimensionalized $p \to p/m_ec$, $v^\prime \to v^\prime/c$, $\sigma_M \to \sigma_M / r^2_e$. Noting that the solution of the adjoint relativistic Fokker-Planck equation does not directly yield the RE distribution $f_e \left( p^\prime, \xi^\prime \right)$, a closure relation will need to be introduced to evaluate the rate of RE generation via avalanching. The simplest closure, introduced in Ref. \cite{Rosenbluth:1997}, involves taking the limit where REs are assumed to have asymptotically high energies and a pitch of $\xi=-1$. While idealized, this closure has been shown to provide a good approximation to the full M\o ller source evaluated using a self-consistently computed RE distribution~\cite{mcdevitt2018relation}.
In the limit of $p^\prime \to \infty$ and $\xi^\prime = -1$, Eq. (\ref{eq:PIE2a}) asymptotes to
\begin{equation}
S \left( p, \xi \right)  = n_e n_{RE} c r^2_e \frac{v}{\gamma^2-1} \frac{1}{\left( \gamma-1 \right)^2} \delta \left( \xi - \xi_1 \right)
, \label{eq:PIE3}
\end{equation}
where we have introduced the Lorentz factor $\gamma \equiv \sqrt{1 + p^2}$ and $\xi_1$ is defined by:
\begin{equation}
\xi_1 = -\sqrt{\frac{\gamma-1}{\gamma+1}}
. \label{eq:PIE4}
\end{equation}
Using Eq. (\ref{eq:PIE3}), Eq. (\ref{eq:PIE1}) reduces to
\begin{equation}
\left. \frac{d n_{RE}}{dt} \right|_{av} = 2\pi n_e n_{RE} c r^2_e \int dp p^2 \frac{v}{\left( \gamma^2-1\right) \left( \gamma - 1 \right)^2} P \left( p, \xi_1 \right)
. \label{eq:PIE6}
\end{equation}
Noting that the right-hand side of Eq. (\ref{eq:PIE6}) is directly proportional to the RE density $n_{RE}$, this implies an exponentially growing solution with a growth rate given by:
\begin{equation}
\tau_c \gamma_{av} = \frac{1}{2\ln \Lambda} \int dp \frac{v}{ \left( \gamma - 1 \right)^2} P \left( p, \xi_1 \right)
, \label{eq:PIE7}
\end{equation}
where $\xi_1$ is defined by Eq. (\ref{eq:PIE4}). Thus, once $P \left( p, \xi \right)$ has been evaluated, the avalanche growth rate can be directly inferred from Eq. (\ref{eq:PIE7}). 

\begin{figure}
\begin{centering}
\includegraphics[scale=0.5]{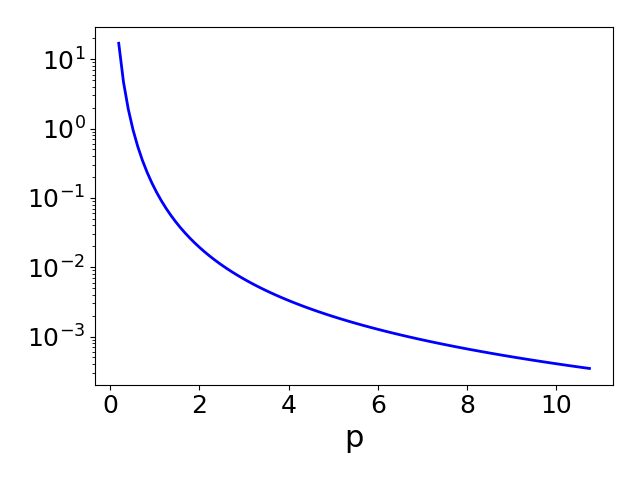}
\par\end{centering}
\caption{The integrand $v / \left( \gamma - 1\right)^2 /\left( 2 \ln\Lambda \right)$ of Eq. (\ref{eq:PIE7}) when $P=1$. The low and high energy bounds are $p_{min} \approx 0.2$ and $p_{max} \approx 10.74$, respectively. The Coulomb logarithm was taken to be $\ln \Lambda = 15$.}
\label{fig:PIE1}
\end{figure}

A caveat when evaluating Eq. (\ref{eq:PIE7}) is that while this integral formally extends to $p\to \infty$, the RPF is evaluated assuming a finite $p_{max}$. The error induced by this approximation can be estimated by considering the magnitude of the integrand in Eq. (\ref{eq:PIE7}) across the range of momenta used in this paper (see Fig. \ref{fig:PIE1}, where $p_{min} \approx 0.2$ and $p_{max} \approx 10.74$). Here we have taken $P=1$ inside the integrand of Eq. (\ref{eq:PIE7}), such that Fig. \ref{fig:PIE1} provides an upper bound on the value of the integrand. Noting that the integrand has decayed to a value of $\approx 3.5\times 10^{-4}$ at the upper boundary, this implies a small contribution to the avalanche growth rate for secondary electrons born with $p > p_{max}$, particularly for parameters where the system is well above marginality. 

\subsection{\label{sec:PDRE}Parametric Dependence of RE avalanche}

\begin{figure}
\begin{centering}
\subfigure[]{\includegraphics[scale=0.5]{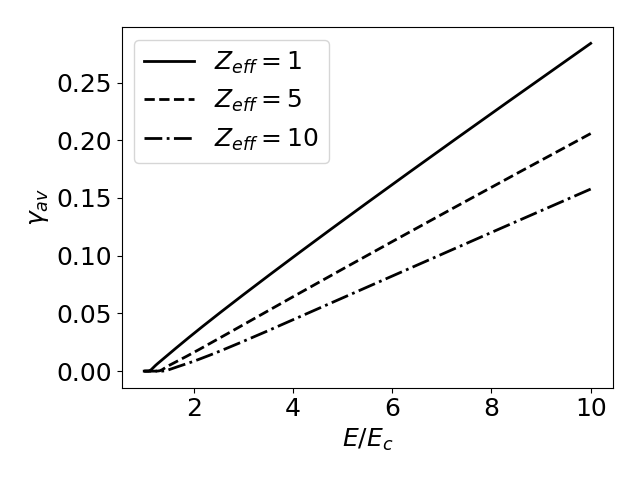}}
\subfigure[]{\includegraphics[scale=0.5]{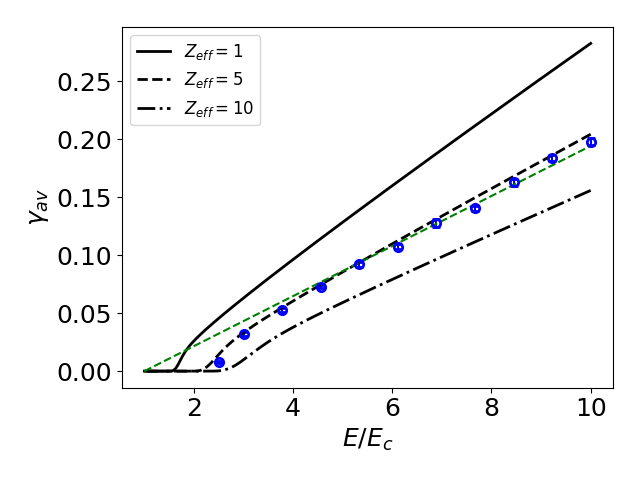}}
\subfigure[]{\includegraphics[scale=0.5]{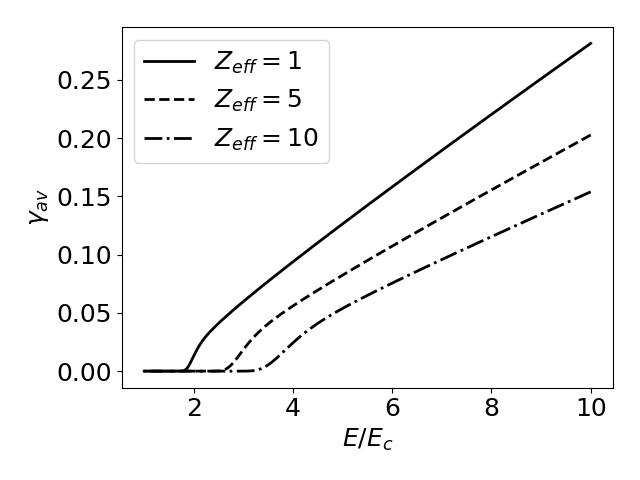}}
\par\end{centering}
\caption{Avalanche growth rate versus electric field for different values of $Z_{eff}$ and $\alpha$. Panel (a) is for $\alpha = 0$, panel (b) is for $\alpha = 0.1$, and panel (c) is for $\alpha = 0.2$. The blue points in panel (b) represent Monte Carlo simulations with $n_e = 5 \times 10^{21}$ m$^{-3}$, and Eq. (\ref{eq:PDRE1}) is shown as the dashed green curve. The Coulomb logarithm was taken to be $\ln \Lambda = 15$.}
\label{fig:PDAM1}
\end{figure}

Using the RPF evaluated in Sec. \ref{sec:RPF}, Eq. (\ref{eq:PIE7}) can be used to infer the avalanche growth rate across the parameter space $\left( E_\Vert, Z_{eff}, \alpha \right)$ (see Fig. \ref{fig:PDAM1}). Here, the avalanche growth rate increases approximately linearly with the electric field when $E_\Vert \gg 1$, with the slope and threshold sensitive to $Z_{eff}$ and $\alpha$. Specifically, the value of $\alpha$ significantly impacts the avalanche threshold (i.e. where $\gamma_{av} \approx 0$), but has a negligible impact at large electric fields. In contrast, $Z_{eff}$ strongly impacts the avalanche growth rate for all values of the electric field. A feature of Eq. (\ref{eq:PIE7}) is that since it indicates the number of REs generated via large-angle collisions, it is thus positive definite. As a result, avalanche growth rates predicted by Eq. (\ref{eq:PIE7}) will not account for the decay of the RE population when $E_\Vert < E_{av}$ and will instead asymptote to zero. While such behavior is strictly correct when focusing solely on the avalanche growth mechanism, when developing a RE model appropriate for coupling with a self-consistent MHD solver it will be necessary to account for the decay of the RE distribution when $E_\Vert < E_{av}$. The inclusion of this additional physics will be the subject of future work.

It will be of interest to compare the predictions of the PINN with available analytic theories. Considering the avalanche growth rate given by Ref. \cite{Rosenbluth:1997}:
\begin{equation}
\tau_c \gamma_{RP} = \frac{1}{\ln \Lambda} \sqrt{\frac{\pi}{2\left( Z_{eff}+5\right)}} \left( E_\Vert - 1 \right)
, \label{eq:PDRE1}
\end{equation}
a comparison between the PINN predictions and Eq. (\ref{eq:PDRE1}) is shown in Fig. \ref{fig:PDAM1}(b). For reference, a small number of avalanche growth rates computed by a Monte Carlo code~\cite{mcdevitt2019avalanche} using the complete M\o ller source are also shown [the blue points on Fig. \ref{fig:PDAM1}(b)]. Details on how this Monte Carlo data set was generated are given in Sec. \ref{sec:PDAM} below. It is apparent that both the PINN and Eq. (\ref{eq:PDRE1}) yield results in reasonable agreement with the Monte Carlo points, where the PINN yields more accurate results near threshold. In particular, while Eq. (\ref{eq:PDRE1}) implies an avalanche threshold field of $E_\Vert = 1$ regardless of the values of $\left( Z_{eff}, \alpha \right)$, the PINN is able to account for the variation of the avalanche threshold for non-zero values of $\alpha$ and different $Z_{eff}$. In addition, at larger electric fields a modest difference in the slope of the avalanche growth rate predicted by Eq. (\ref{eq:PDRE1}) compared to the Monte Carlo data is evident.
The predictions of the PINN, in contrast, are able to recover the correct slope of the avalanche growth rate, albeit with a modest offset in the magnitude as will be discussed further in Sec. \ref{sec:PDAM} below.

\begin{figure}
\begin{centering}
\subfigure[]{\includegraphics[scale=0.5]{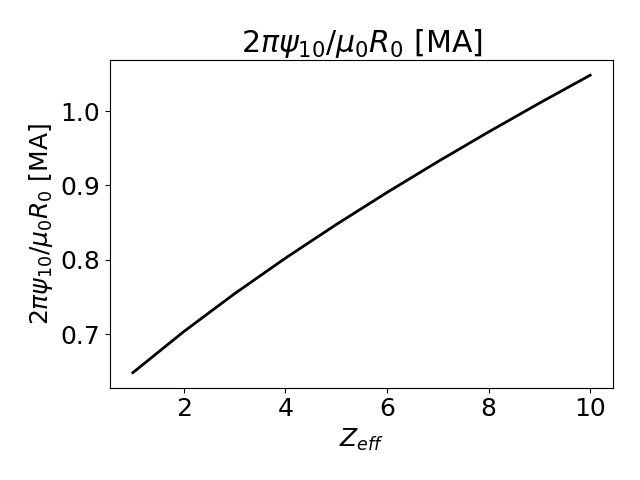}}
\subfigure[]{\includegraphics[width=0.5\textwidth]{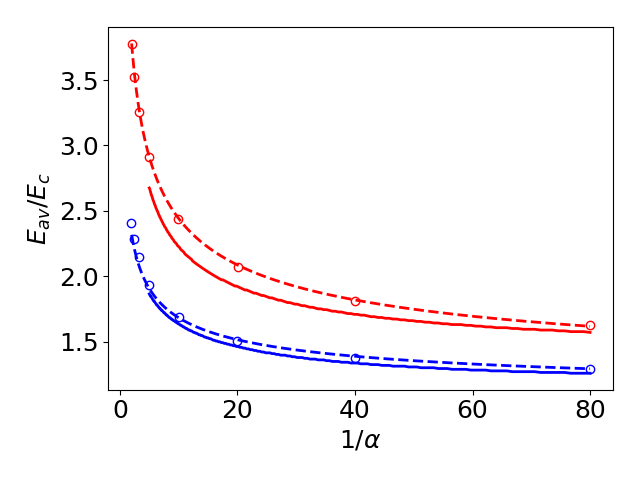}}
\par\end{centering}
\caption{(a) The value of $2\pi \psi_{10}/\mu_0 R_0$ as a function of $Z_{eff}$, with a Coulomb logarithm of $\ln \Lambda = 15$. (b) Avalanche threshold $E_{av}$ as a function of the synchrotron radiation strength $\alpha$. The solid lines represent the PINN predictions, the dashed lines and `o' markers represent Eq. (B15) and Monte Carlo results from Ref.~\cite{mcdevitt2019avalanche}, respectively, and WebPlotDigitizer was used to extract the Monte Carlo values. The blue values represent $Z_{eff} = 1$ and the red values represent $Z_{eff} = 5$. The Coulomb logarithm was taken to be $\ln\Lambda = 20$.}
\label{fig:PDAM2}
\end{figure}

When well above marginality the avalanche growth rate is most conveniently described by evaluating the amount of poloidal flux needed to increase the amplitude of a RE seed by one order of magnitude~\cite{Rosenbluth:1997, Boozer:2018}. This quantity can be evaluated by noting that for $E_\Vert \gg 1$, the avalanche growth rate scales linearly with $E_\Vert$, i.e.
\begin{equation}
\gamma_{av} = \gamma_0 \left( \frac{E_\Vert}{E_c} - 1\right) \approx \gamma_0 \frac{E_\Vert}{E_c} \approx \frac{\gamma_0}{E_c} \frac{1}{R_0} \frac{\partial \psi}{\partial t}
, \label{eq:PDAM1}
\end{equation}
where $\gamma_0$ is a constant that depends on $\left( Z_{eff}, \alpha, \ln \Lambda \right)$, $\ln \Lambda$ is the Coulomb logarithm, $R_0$ is the major radius, and $\psi$ is the poloidal flux function. After integrating Eq. (\ref{eq:PDAM1}) over the time interval $t_f - t_i$, the number of exponentials of the RE population can be written in terms of the change of poloidal flux, and a constant $\psi_{\text{exp}}$ that defines the efficiency of the avalanche. In particular, integrating Eq. (\ref{eq:PDAM1}) over the time interval $t_f - t_i$, yields
\begin{equation}
N_{exp} = \int^{t_f}_{t_i} dt \gamma_{av} \approx \frac{\gamma_0}{E_c} \frac{1}{R_0} \int^{t_f}_{t_i} dt \frac{\partial \psi}{\partial t} = \frac{\Delta \psi}{ \psi_{\text{exp}}}
, \label{eq:PDAM2}
\end{equation}
where $\Delta \psi \equiv \psi \left( t_f\right) - \psi \left( t_i\right)$ and we have defined $\psi_{exp} \equiv R_0 E_c/ \gamma_0$, which is related to the amount of poloidal flux required to effect one exponential amplification of the RE population. Thus, once $\gamma_0$ is inferred, the efficiency through which the decay of the poloidal flux leads to an amplification of the seed RE population can be evaluated. Further defining the quantity $\psi_{10}\equiv \ln 10 \psi_{exp}$ (which delineates base ten amplifications), the efficiency of the avalanche growth rate for a broad range of parameters is shown in Fig. \ref{fig:PDAM2}(a). Here, the avalanche growth rate is most efficient for low values of $Z_{eff}$ and increases by nearly a factor of two for $Z_{eff}=10$, agreeing with previous results \cite{mcdevitt2019avalanche}. We have not indicated the dependence of $\psi_{10}$ on $\alpha$ since for large electric fields, the avalanche growth rate will be independent of $\alpha$. 

In addition, the avalanche threshold is strongly impacted by the parameters $\left( Z_{eff}, \alpha \right)$. In particular, as $Z_{eff}$ or $\alpha$ are increased, the threshold electric field $E_{av}$, where the avalanche growth rate is zero, is increased. The dependence of the avalanche threshold on $\left( Z_{eff}, \alpha \right)$ is shown in Fig. \ref{fig:PDAM2}(b), where the PINN predictions are the solid lines. The Monte Carlo results are the `o' markers, and the dashed lines represent empirical fit formula given by Eq. (B15) of Ref.~\cite{mcdevitt2019avalanche}. Here, since the avalanche growth rates predicted by the PINN are positive definite, we define the avalanche threshold as the value of the electric field where $\tau_c \gamma_{av} = 2\times 10^{-3}$ . From Fig. \ref{fig:PDAM2}(b) it is apparent that the PINN predicts $E_{av}$ particularly well across $\alpha$ for $Z_{eff} = 1$, with somewhat larger deviations evident for $Z_{eff}=5$. The systematic under-prediction of the avalanche threshold is due to the assumption in Eq. (\ref{eq:PIE7}) that primary electrons have infinite energy. Such an assumption overestimates the number of secondary electrons generated, and perhaps more importantly, neglects that the primary electron population itself will slowly decay to the thermal bulk when $E \approx E_{av}$.

\subsection{\label{sec:PDAM}Verification of the Surrogate Model}

In this section we will verify the PINN's predictions of the avalanche growth rate against first principle Monte Carlo simulations across the training region ($E_\Vert$, $Z_{eff}$, $\alpha$). The RunAway Monte Carlo (RAMc) code is employed (see Ref. \cite{mcdevitt2019avalanche} for a detailed description), which evolves the guiding center motion of relativistic electrons and includes effects from small-angle collisions, large-angle collisions, and synchrotron radiation. Large-angle collisions are evaluated using the full M\o ller source~\cite{Moller:1932}. In order to avoid toroidal corrections to the avalanche growth rate~\cite{mcdevitt2019runaway, arnaud2024impact}, all REs are initialized near $r=0$, and a large tokamak device was chosen with a minor radius of $a=2\;\text{m}$ and major radius $R_0=6\;\text{m}$, in order to render spatial transport negligible. A geometry with circular flux surfaces was selected, with a safety factor profile taken to be $q \left( r \right) = 2.1 + 2\left( r/a\right)^2$.

\begin{figure}
\begin{centering}
\subfigure[]{\includegraphics[scale=0.25]{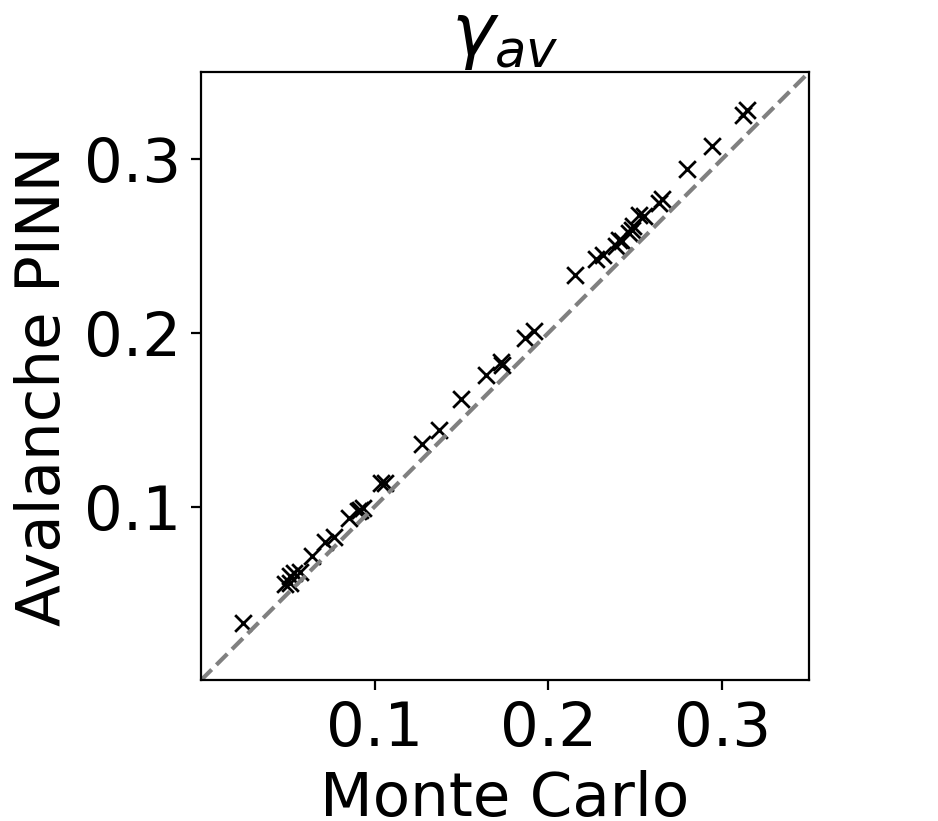}}%
\subfigure[]{\includegraphics[scale=0.25]{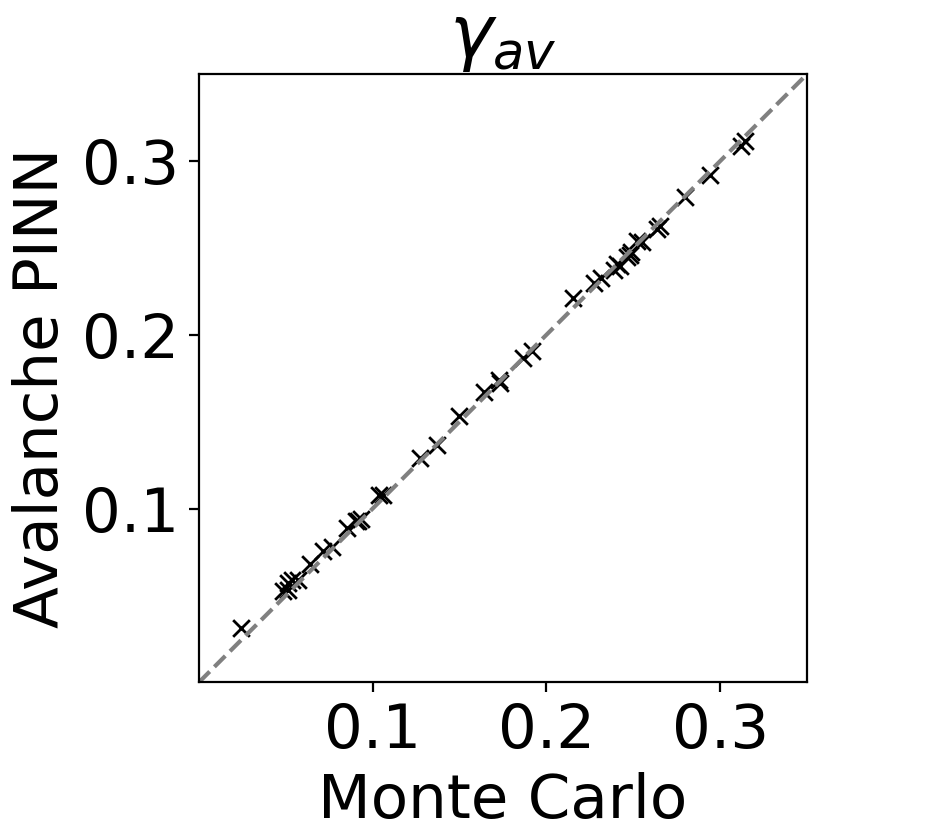}}
\subfigure[]{\includegraphics[scale=0.25]{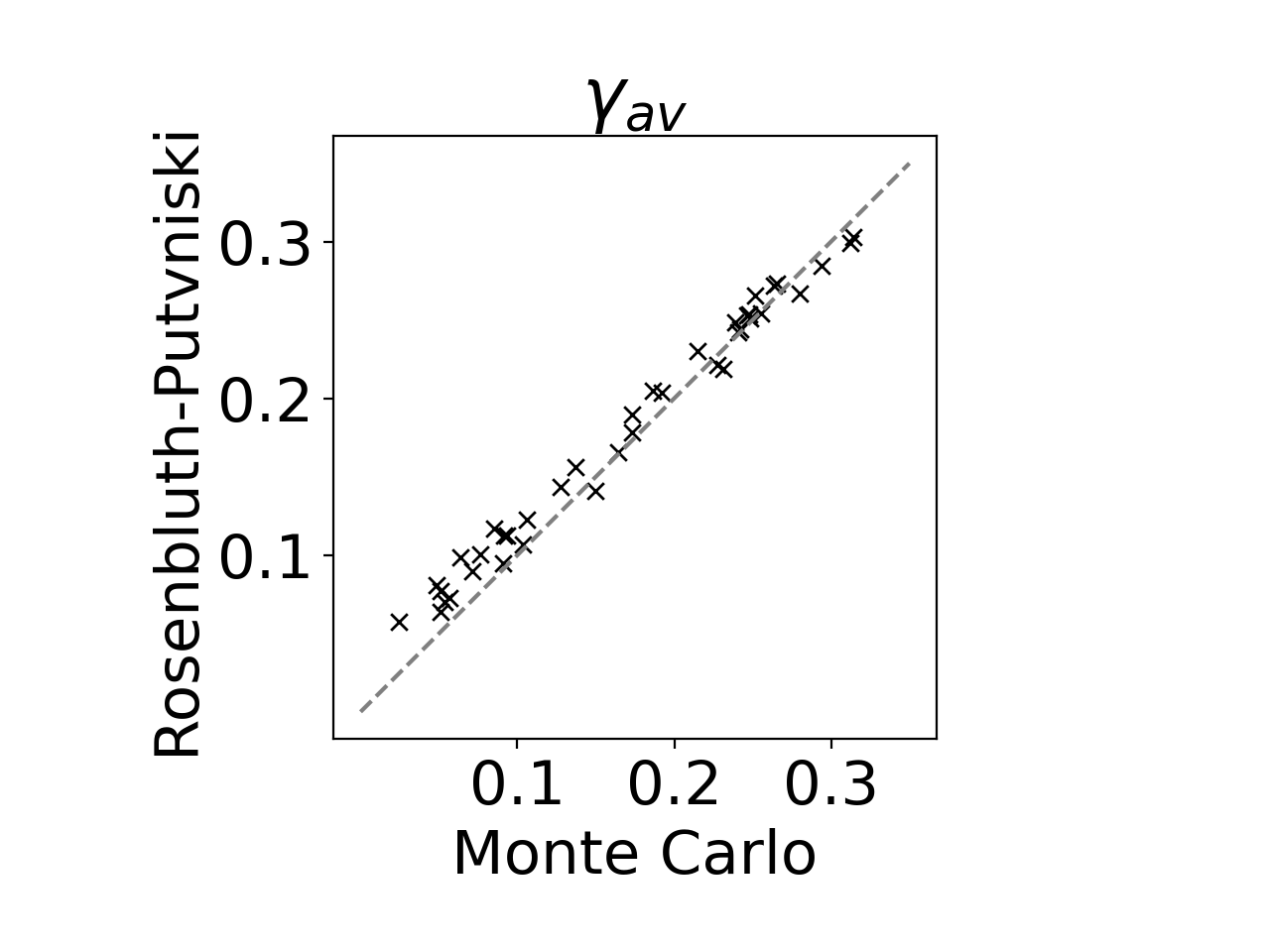}}
\par\end{centering}
\caption{(a) Avalanche growth rate comparison between the PINN and Monte Carlo solver. The grey dashed line represents the coefficient of determination of $R^2 = 1$.  (b) Same as panel (a), but with the predictions of the avalanche PINN multiplied by the factor $0.94886$. (c) Avalanche growth rate comparison between Eq. (\ref{eq:PDRE1}) and the Monte Carlo solver. The avalanche growth rates were evaluated across $E_\Vert \in$ (1,10), $Z_{eff} \in$ (1,10), and $\alpha \in \left(2.8 \times 10^{-3},0.2\right)$. The other parameters were chosen to be $T_e$ = 10 eV and $n_e$ = 10$^{21}$ m$^{-3}$.}
\label{fig:Comparison}
\end{figure}

The Monte Carlo avalanche simulations are set up by initializing a small population of electrons (8 for this analysis) at high momentum [$p \in$ (10,20)] and strongly aligned with the magnetic field [$\xi \in$ (-0.9,-1.0)]. They are then allowed to exponentially grow in time until a saturated growth rate can be identified, generally resulting in the initial seed RE population growing by several orders of magnitude. The plasma parameters chosen were a density of $n_e = 10^{21}$ m$^{-3}$, a temperature of $T_e = 10$ eV, and a toroidal magnetic field strength that was varied to give the appropriate value of $\alpha$. One caveat is that at a fixed density and temperature $\alpha \propto B^2$, which in turn can lead to orbit drifts at the low boundary of synchrotron radiation due to the weak magnetic field. These orbit drifts are thus ensured to be negligible by enforcing a minimum toroidal magnetic field of $B = $2 T, which corresponds to a lower bound ($\alpha \approx 2.8 \times 10^{-3}$) in the training region for synchrotron radiation. Fifty randomly selected samples of ($E_\Vert$, $Z_{eff}$, $\alpha$) are chosen as input parameters for the Monte Carlo solver and the avalanche PINN, where cases below the avalanche threshold are discarded, leaving forty-one samples. A comparison between the predictions of the avalanche PINN and Monte Carlo solver is shown in Fig. \ref{fig:Comparison}(a). Here, the grey dashed line represents equality (y-axis equal to x-axis) between the PINN and Monte Carlo predictions. The coefficient of determination~\cite{guilford1936personality} between the PINN predictions and the Monte Carlo avalanche growth rate was evaluated to be $R^2 \approx 0.9849$, indicating that the PINN was able to accurately predict the avalanche growth rate across a broad range of parameters. One noticeable feature present in Fig. \ref{fig:Comparison}(a) is the systematic over prediction of the avalanche growth rate by the PINN. This feature is expected, and is due to the use of the simplified Rosenbluth-Putvinski secondary source term described by Eq. (\ref{eq:PIE3}), which assumes all primary electrons to have $p \to \infty$ and $\xi=-1$. We note that this feature can be mitigated by introducing an order unity factor to correct for the discrepancy between the PINN and Monte Carlo predictions. After optimization, it was found that multiplying the predictions of the PINN by a factor of 0.94886 significantly reduced the discrepancy between the PINN and Monte Carlo predictions. After applying this factor [see Fig. \ref{fig:Comparison}(b)], the resulting coefficient of determination increases to $R^2$ = 0.9988. 

The PINN is also shown to robustly provide better avalanche growth rate predictions than the analytical expression provided by Eq. (\ref{eq:PDRE1}) [compare Figs. \ref{fig:Comparison}(a) and (c)]. Here, the avalanche growth rate provided by Eq. (\ref{eq:PDRE1}) compared with the Monte Carlo solver has a weaker correlation than that between the PINN and Monte Carlo solver, where the coefficient of determination for Fig. \ref{fig:Comparison}(c) was $R^2 \approx 0.9644$. The avalanche PINN thus improves on the accuracy of the avalanche growth rate predictions, even in the absence of the 0.94886 factor.

\section{\label{sec:SC}Summary and Conclusions}

This work utilized a physics-informed neural network to evaluate the steady state solution of the adjoint to the relativistic Fokker-Planck equation.
Noting that the PINN takes the physical parameters of the problem as inputs, once trained the PINN provides an efficient surrogate model of the RPF. 
In addition, while a comprehensive description of the avalanche growth rate requires evaluating the primary electron distribution, a quantity not evaluated in the present approach, by invoking the often employed simplification that primary electrons have infinite energy and a pitch $\xi=-1$~\cite{Rosenbluth:1997} the RE avalanche growth rate was shown to be directly linked to an integral of the RPF [Eq. (\ref{eq:PIE7})]. While this approximation to the secondary source of REs is known to lead to a modest overestimate of the avalanche growth rate at modest values of the electric field~\cite{mcdevitt2019avalanche}, predictions from the PINN were shown to agree with direct Monte Carlo simulations that utilized a complete M\o ller source across a broad range of parameters with an offset of roughly 5\% [Fig. \ref{fig:Comparison}(a)]. This small offset can be largely removed by multiplying predictions of the PINN by a factor of order unity [compare Figs. \ref{fig:Comparison}(a) and \ref{fig:Comparison}(b)]. We note, however, that such a correction will have little impact on the avalanche threshold, and thus the present model predicts a smaller avalanche threshold electric field compared to predictions using a full M\o ller source [see Fig. \ref{fig:PDAM2}(b)], though it does recover the dominant dependencies on the physics parameters $\left( Z_{eff}, \alpha \right)$.

An additional aim of this paper was to provide a proof-of-principle demonstration that physics constrained deep learning methods offer an attractive avenue through which RE surrogate models can be developed. In contrast to data driven deep learning approaches, the present method encodes the underlying kinetic solution into the neural network, rather than just the QoI (avalanche growth rate in this case), and thus allows for greater interpretability and hence greater confidence in the accuracy of the prediction. While the present paper has focused on an idealized description of REs, generalization to more complete models of RE formation, incorporating partial screening effects~\cite{Hesslow:2017} for example, can be accomplished by modifying the collision coefficients in the adjoint equation described by Eq. (\ref{eq:LDCF2}) and retraining the PINN. This extension will be the subject of future work. We thus anticipate that the present approach provides a flexible means through which RE surrogate models can be developed for a broad range of plasma conditions.

\begin{acknowledgements}

This work was supported by DOE OFES under award Nos. DE-SC0024634 and DE-SC0024649. The authors acknowledge the University of Florida Research Computing for providing computational resources that have contributed to the research results reported in this publication.

\end{acknowledgements}

\newpage

\bibliographystyle{apsrev}
\bibliography{../../bib_files/ref}

\end{document}